\documentclass[aps,prl,twocolumn,superscriptaddress,longbibliography,preprintnumbers]{revtex4-1}

\usepackage{mathrsfs}
\usepackage{slashed}
\usepackage{amsmath}
\usepackage{amsfonts}
\usepackage{graphicx,color}
\usepackage{times}
\usepackage[caption=false]{subfig}
\usepackage[colorlinks,linkcolor=red,citecolor=blue]{hyperref}
\usepackage{braket}

\AtBeginDocument{%
    \newwrite\bibnotes
    \def\bibnotesext{Notes.bib}
    \immediate\openout\bibnotes=\jobname\bibnotesext
    \immediate\write\bibnotes{@CONTROL{REVTEX41Control}}
    \immediate\write\bibnotes{@CONTROL{%
    apsrev41Control,author="08",editor="1",pages="1",title="0",year="1"}}
     \if@filesw
     \immediate\write\@auxout{\string\citation{apsrev41Control}}%
    \fi
}%

\begin{document}

\title{Exceptional dynamics of interacting spin liquids}
\author{Kang Yang}
\author{Daniel Varjas}
\author{Emil J. Bergholtz}
\affiliation{Department of Physics, Stockholm University, AlbaNova University Center, 106 91 Stockholm, Sweden}

\author{Sid Morampudi}
\affiliation{Center for Theoretical Physics, MIT, Cambridge MA 02139, USA}

\author{Frank Wilczek}
\affiliation{Department of Physics, Stockholm University, AlbaNova University Center, 106 91 Stockholm, Sweden}
\affiliation{Center for Theoretical Physics, MIT, Cambridge MA 02139, USA}
\affiliation{T. D. Lee Institute, Shanghai, China} 
\affiliation{Wilczek Quantum Center, Department of Physics and Astronomy, Shanghai Jiao Tong University, Shanghai 200240, China}
\affiliation{Department of Physics and Origins Project, Arizona State University, Tempe AZ 25287 USA}

\begin{abstract}
 We show that interactions in quantum spin liquids can result in non-Hermitian phenomenology that differs qualitatively from mean-field expectations. We demonstrate this in two prominent cases through the effects of phonons and disorder on a Kitaev honeycomb model. Using analytic and numerical calculations, we show the generic appearance of exceptional points and rings depending on the symmetry of the system. Their existence is reflected in dynamical observables including the dynamic structure function measured in neutron scattering. The results point to new phenomenological features in realizable spin liquids that must be incorporated into the analysis of experimental data and also indicate that spin liquids could be generically stable to wider classes of perturbations.  
\end{abstract}
\preprint{MIT-CTP/5402}
\maketitle


Quantum spin liquids are low temperature phases of matter in which quantum fluctuations prevent the establishment of long-range magnetic order. Besides the absence of local order, a more distinct characteristic is the presence of exotic fractionalized spin excitations (spinons) and emergent gauge fields~\cite{anderson1987resonating, balents2010spin, savary2016quantum, Knolle2019} due to long-range entanglement in the system. This suggests possible applications ranging from quantum simulation to spintronics~\cite{yang2021probing}.  

Much recent work has focused on realizing new types of spin liquids and understanding their implications in dynamics and experiments through mean-field approaches. Interactions and disorder are prominent in many experimental settings, however, and can affect the dynamics and thermodynamics~\cite{sibille2017coulomb, kimchi2018valence, han2016correlated, cao2016low, zhu2017disorder, li2017crystalline, pasco2018single, lee2009low, mross2010controlled, lee2005u, morampudi2017statistics, morampudi2020spectroscopy, pace2021emergent}. A common expectation is that such effects either renormalize the properties of quasiparticles (and give them finite lifetimes), or open a gap if they violate certain symmetries. Here, we explore an intriguing alternative route where interactions and disorder can generically lead to qualitatively new phenomena, through distinct non-Hermitian effects that depend on the symmetries of the interactions \cite{RevModPhys.93.015005,PhysRevX.8.031079,PhysRevLett.121.026403,PhysRevLett.125.227204,PhysRevB.99.201107,PhysRevB.100.245205,PhysRevB.99.165145,yoshida2020exceptional,PhysRevB.104.035413,PhysRevB.104.L121109,PhysRevB.101.085122}. These non-Hermitian components can induce a unique level attraction in contrast with usual band degeneracies of Hermitian perturbations  We will illustrate this general principle in the context of the Kitaev honeycomb model, which has received much attention recently in view of potential experimental realizations~\cite{jackeli2009mott, takagi2019concept, banerjee2017neutron, kitagawa2018spin, kasahara2018majorana, vinkler2018approximately, ye2018quantization, rau2016spin, gohlke2018quantum, hermanns2018}, whereas our symmetry analysis may apply to a variety of models. The presence of disorder and phonons can lead to appearance of so-called exceptional rings and exceptional points which possess an unusual square-root dispersion \cite{Berry2004,PhysRevLett.126.077201}. This results in unusual features in experimental observables including asymmetric Fermi arcs, which cannot be achieved generically in Hermitian settings. The resulting generic phenomena also illustrate that spin liquids can be stable to a wider variety of perturbations and are less fragile and richer than typically assumed.

\begin{figure}
    \centering
    \includegraphics[width=1.0\linewidth]{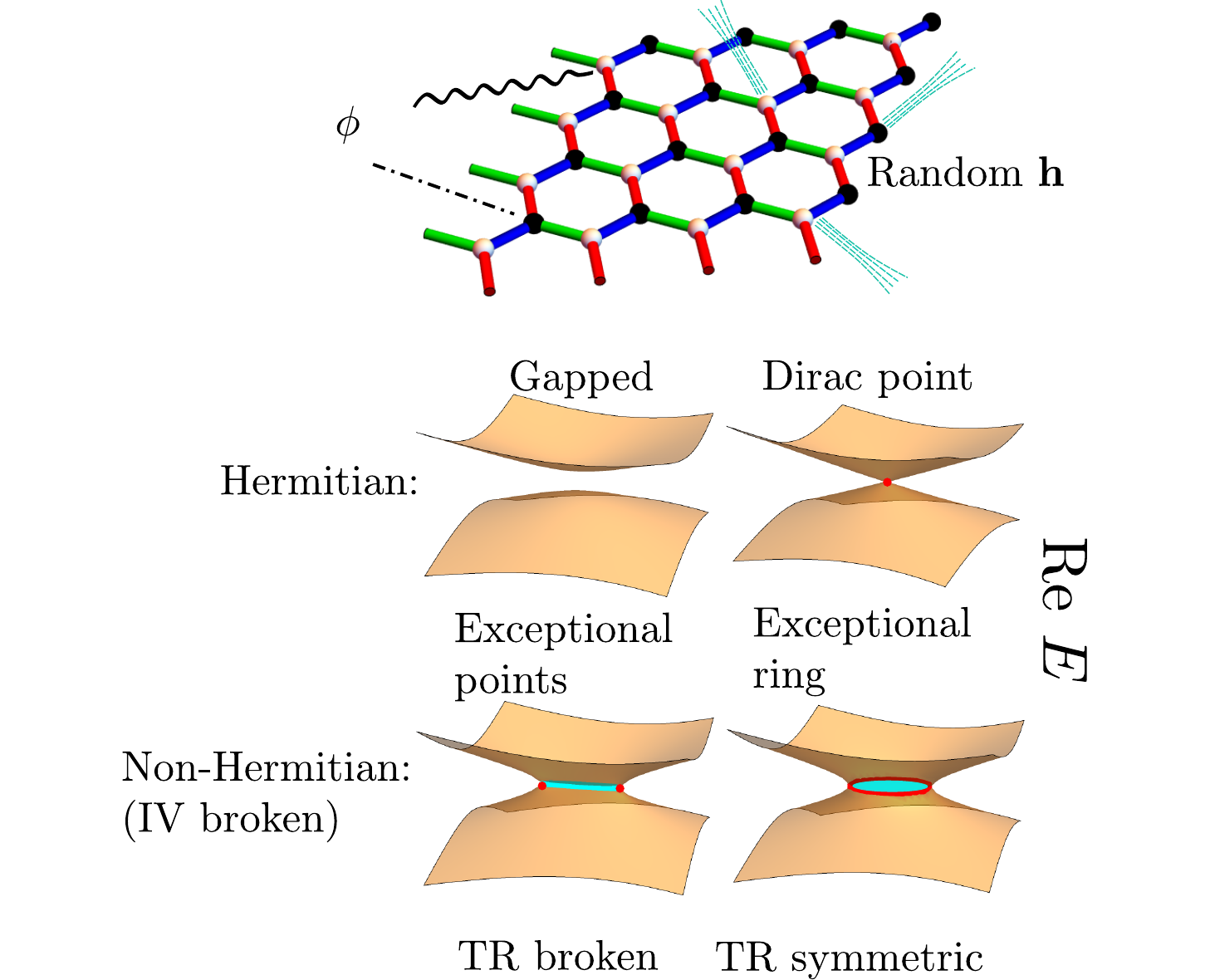}
    \caption{The Kitaev honeycomb model with interactions carried by a bosonic field or bond disorders. Depending on whether inversion (IV) symmetry and time-reversal (TR) symmetry are preserved, we can have different types of non-Hermitian phases or a trivial gapped phase. The real part of the "energy" difference $\Delta E$ (yellow) between the two Majorana bands are shown as a function of momenta for different scenarios.}
    \label{fig_sm}
\end{figure}

\renewcommand{\arraystretch}{1.5}
\begin{table*}[t]
\centering
\begin{tabular}{ |p{5.9cm}|p{4.3cm}|p{5.5cm}|  }
\hline
\multicolumn{3}{|c|}{Symmetry of Green's functions} \\
\hline
Time reversal & Particle-hole& Inversion \\
\hline
$\mathcal T\psi_{a,\mathbf k}\mathcal T^{-1}=\sum_b U_{ab}\psi_{b,\mathbf k}$ & $\mathcal C\psi_{a,\mathbf k}\mathcal C^{-1}=\psi^\dagger_{a,-\mathbf k}$ & $\mathcal P\psi_{a,\mathbf k}\mathcal P^{-1}=\sum_b U^I_{ab}\psi_{b,-\mathbf k}$ \\
\hline
$\Sigma^R_{ab}(\omega,\mathbf k)=\sum_{cd}U_{ac}U^\ast_{bd}\left[\Sigma^A_{cd}(\omega,-\mathbf k)\right]^\ast$ & $\Sigma^R_{ab}(\omega,\mathbf k)=-\Sigma^A_{ba}(-\omega,-\mathbf k)$ & $\Sigma^R_{ab}(\omega,\mathbf k)=\sum_{cd}U^I_{ac}U^{I\ast}_{bd}\Sigma^R_{cd}(\omega,-\mathbf k)$ \\
\hline
$E(\mathbf k)\to E(-\mathbf k)$ &$E(\mathbf k)\to -E^\ast(-\mathbf k)$& $E(\mathbf k)\to E(-\mathbf k)$\\
\hline
\end{tabular}
\caption{Summary of symmetries. We use a general complex fermion notion. For Majorana operators used here, we need to substitute $\psi_{a,\mathbf k}=c_{a,\mathbf k},\psi^\dagger_{a,\mathbf k}=c_{a,-\mathbf k}$.}
\label{tb_tsfe}
\end{table*}

\paragraph*{Effective non-Hermitian description}

Elucidating the subtle signatures of a spin liquid in experiments requires an understanding of dynamical observables such as the spectral function and dynamic structure factor. Linear response connects these observables to tangible measurements such as scattering cross-sections. Calculating these observables in interacting systems is not easy, although they have been quantified in some crucial cases such as an interacting electron gas (Fermi liquid theory). 

A fundamental object to calculate dynamical observables is the Greens function where interactions are accounted for through a self-energy.
The Green's function for an interacting or disordered system satisfies the Dyson equation $[\omega-H_0-\Sigma^{A/R}(\omega)]G^{A/R}(\omega)=I$, where $I$ is the identity operator and  $H_0$ is the unperturbed Hamiltonian, and the superscripts $A,R$ refer to advanced or retarded. The retarded version is appropriate for calculating the time evolution of simply specified initial (``in'') states.  The self-energy $\Sigma^{A/R}(\omega)$ terms are induced by the interaction or the disorder. The Green's function has poles at $\omega=E$ whenever $\det[E-H_0-\Sigma^{A/R}(E)]=0$. At low energy, the self-energy can be expanded in powers of $\omega$, $\Sigma^{R}(\omega)=\Sigma^{R/A}(0)+\omega \Sigma^{(1)} +\dots$. Unlike the original Hamiltonian, the self-energy is usually not Hermitian \cite{peskin2018introduction}. We denote the Hermitian and anti-Hermitian component of the self-energy as $\bar \Sigma$ and $\tilde \Sigma$. The linear term $\Sigma^{(1)}$ plays an important role in wave function renormalization and friction for bosonic operators \cite{wen2007quantum}. Here we assume $||\Sigma^{(1)}||\ll 1$ and neglect it for simplicity. The leading-order term in the Dyson equation can be treated as an effective Hamiltonian $H^{\textrm{eff}}=H+\Sigma^{R}(0)$.

\paragraph*{Symmetries of interactions and exceptional degeneracies}
This effective Hamiltonian is usually non-Hermitian and can exhibit "exceptional" degeneracies in its spectrum depending on the symmetries obeyed by the interactions \cite{PhysRevX.9.041015,PhysRevB.99.041406,PhysRevB.99.121101}. We illustrate this through the Kitaev honeycomb model.

The Kitaev honeycomb model \cite{KITAEVHoneycomb} is defined through compass interactions linking directions in spin space and real space of spin-$1/2$:
\begin{equation}
    H_0=-\sum_{\langle jk\rangle_\alpha} J_\alpha\sigma^\alpha_j\sigma^\alpha_k,\label{eq_oHam}
\end{equation}
where $\langle jk\rangle_\alpha$ labels the lattice (Fig.~\ref{fig_sm}) and $\alpha=x,y,z$ labelling the three types of links of a hexagonal lattice with $\sigma^\alpha$ the corresponding Pauli matrices. At low energies, the system is effectively described by Majorana quasiparticles $c_a$ with a Dirac dispersion interacting with a $\mathbb{Z}_2$ gauge field, where $a$ labels the two sublattice indices.

The Majorana operators always possess a particle-hole symmetry, $Cc_{a,\mathbf k}C^{-1}=c_{a,-\mathbf k}$, reflecting the underlying real bosonic spin degrees of freedom. The honeycomb lattice is inversion-invariant: under inversion transformation, the two orbitals are interchanged while the $\mathbb Z_2$ gauge field coupling the two orbitals obtains a minus sign. Therefore the Majorana operators transform as $\mathcal Pc_{a,\mathbf k}\mathcal P^{-1}=\sum_b\varepsilon_{ab}c_{b,-\mathbf k}$. The time-reversal symmetry is crucial in protecting the gapless phase. The transformation rule is given by $\mathcal Tc_{1,\mathbf k}\mathcal T^{-1}=c_{1,-\mathbf k}, \mathcal Tc_{2,\mathbf k}\mathcal T^{-1}=-c_{2,-\mathbf k}$. With these rules, we can explicitly compute how the self-energy transforms under $\mathcal T,\mathcal P$ and $\mathcal C$. They are summarized in Table~\ref{tb_tsfe}.

From these symmetries, we find different types of exceptional degeneracies. We first look at PH symmetry and inversion symmetry. PH symmetry requires the NH component of the self-energy to satisfy $\tilde\Sigma_{ab}^{R}(\omega,\mathbf k)=\tilde\Sigma^R_{ba}(-\omega,-\mathbf k)$. Inversion symmetry imposes $\tilde\Sigma^R_{22}(\omega,\mathbf k)=\tilde\Sigma^R_{11}(\omega,-\mathbf k)$ and $\tilde\Sigma^R_{12}(\omega,\mathbf k)=-\tilde\Sigma^R_{21}(\omega,-\mathbf k)$. So when both of them are present, the NH self-energy can only be proportional to an identity matrix at $\omega=0$, $\tilde \Sigma^R_{ab}(0,\mathbf k)
=\delta_{ab}\tilde\Sigma^R(0,\mathbf k)$. Under this circumstance, the NH components is trivial, merely broadening the resonance peaks of the Majorana operators. So in order for a non-trivial NH self-energy, we need to break either PH or inversion symmetry. The former is an intrinsic property of Majorana operators. Thus we can only choose to break the inversion symmetry.

Looking at the time-reversal symmetry, it requires $\tau_z\Sigma^R(\omega,\mathbf k)\tau_z=[\Sigma^A(\omega,-\mathbf k)]^\ast$ with $\tau_z=\textrm{diag}(1,-1)$ the $z$-Pauli matrix. Together with PH symmetry, we find that time-reversal symmetry implies $\bar\Sigma^R_{aa}(0,\mathbf k)=0$ and $\tilde\Sigma^R_{ab}(0,\mathbf k)=0(a\ne b)$. We can only have purely imaginary numbers in the diagonals of $\Sigma^R(0,\mathbf k)$ when time-reversal symmetry is present. 

With the symmetry restrictions, we discuss the topology of band touchings. The Majorana Hamiltonian can be expressed in terms of the Pauli matrices $H_m=d_0I+\mathbf d\cdot\boldsymbol{\tau}$. The $d_0$ vector merely shifts the touching energy level and we only need to focus on $E=\pm\sqrt{\mathbf d\cdot\mathbf d}$. When the system preserves time-reversal symmetry, we have $d_z=i\tilde d$ and $d_{x/y}=\bar d_{x/y}$. So we have 
\begin{equation}
    E(\mathbf k)=\pm \sqrt{\bar d^2_x(\mathbf k)+ \bar d^2_y(\mathbf k)-\tilde d^2_z(\mathbf k)}.
\end{equation}
Now the energy vanishes at $\bar d^2_x(\mathbf k)+ \bar d^2_y(\mathbf k)=\tilde d^2_z(\mathbf k)$ and is purely imaginary when $\bar d^2_x(\mathbf k)+ \bar d^2_y(\mathbf k)<\tilde d^2_z(\mathbf k)$. The solution to  $\bar d^2_x(\mathbf k)+ \bar d^2_y(\mathbf k)=\tilde d^2_z(\mathbf k)$ is a $1$D closed curve, the exceptional curve. Inside the exceptional curve, the energy is purely imaginary; the system is gapless for the real part of the energy. When the time-reversal symmetry is lifted, the system usually possesses a gap $\Delta\sim\textrm{min} \sqrt{\bar d^2(\mathbf k)}$. However, the inclusion of a nontrivial NH term can change the situation. The energy is given by
\begin{equation}
     E(\mathbf k)=\pm \sqrt{\bar d^2(\mathbf k)-\tilde d^2(\mathbf k)+2i\bar {\mathbf d}(\mathbf k)\cdot\tilde{\mathbf d}(\mathbf k)}.
\end{equation}
By dimension counting, the imaginary and real parts in the square root can vanish robustly at an isolated point $\mathbf k=\mathbf k_\ast$ since these are two restrictions for a $2$D problem. This is different from the Hermitian case where all the three components $d_x,d_y$ and $d_z$ must vanish. The NH components generate a level attraction and may close the (real) gap, leading to an exceptional point at $\mathbf k_\ast$. 

Exceptional points are general features of a NH effective Hamiltonians. However, the exceptional point itself is not directly reflected in the single-body spectral function. The existence of exceptional points is always accompanied by  Fermi arcs. Compared to the Dirac point, the Fermi-arc is extensive in one direction while narrow in the orthogonal direction. Its effective dispersion is also biased at finite frequency (see \cite{suppm}). Such a highly anisotropic feature could be observed in inelastic neutron scattering. 

\paragraph*{Exceptional rings in a disordered Kitaev honeycomb model}

\begin{figure}[t]
\includegraphics[width=1\linewidth]{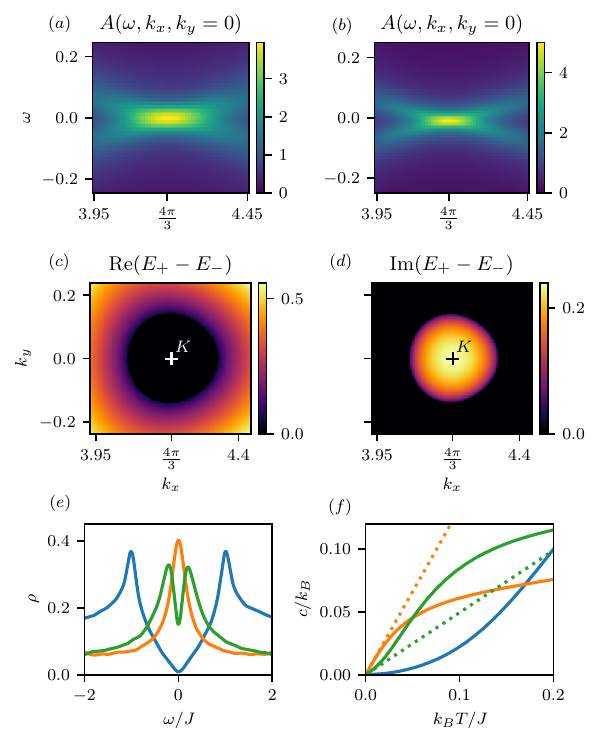}
\caption{Exceptional ring in a disordered Kitaev honeycomb model. Spectral function along the $k_y = 0$ cut from disordered numerics $(a)$ and SCBA $(b)$.
Real $(c)$ and imaginary $(d)$ part of the gap in the effective Hamiltonian near the $K$ point.
The real gap vanishes inside, the imaginary gap outside the exceptional ring.
$(e)$ Density of states and $(f)$ specific heat for clean (blue), nearest-neighbor disordered without exceptional ring (orange), and second-neighbor disordered with exceptional ring (green) systems, the low-temperature linearized specific heat with dotted lines.}
\label{fig:ring}
\end{figure}

\begin{figure*}[t]
\includegraphics[width=1.0\linewidth]{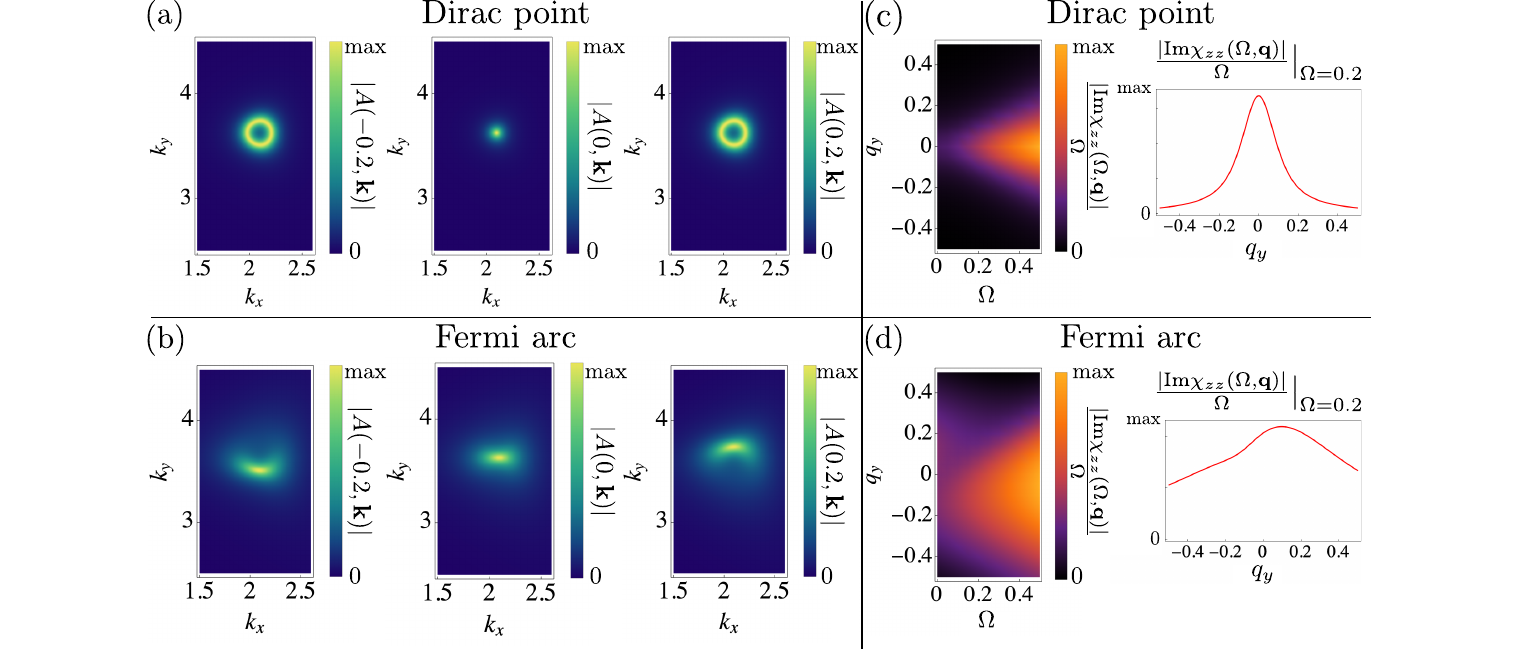} 
\caption{The spectral functions of Majorana excitations  $A(\omega,\mathbf k)=-\textrm{tr}[G^R(\omega,\mathbf k)-G^{R\dagger}(\omega,\mathbf k)]/(2i\pi)$, and their spin structure factors in the presence of Heisenberg and cross term interactions. The energy is unit is $J$. (a)-(c) The Dirac point for Hermitian systems. The single-particle spectral function is supported at a point. The spin structure factor as a function of the frequency $\Omega$ and momentum $\mathbf q$ takes a conic structure for both the $\mathbf q_x$-direction cut and the $\mathbf q_y$-direction cut. (c)-(f) A Fermi arc with NH components $\tilde d_0=-0.35J, \tilde d_x=0.25J$, associated with the exceptional points in the single-body Majorana spectral function. The spin structure is now anisotropic due to the strong asymmetric behaviours of the Fermi arc. (g)-(h) The spectral functions of the Fermi arc at finite positive and negative frequencies. The Fermi arc becomes convex at finite frequencies. The direction of the deformation depends on the sign of the frequency. (i) The cuts of the spin structure factor at $\Omega=0.2J$ along $(0,q_y)$ and $(q_x,0)$.} 
\label{fig_DPFR}
\end{figure*}

Now we specialize to the Kitaev honeycomb model, a
$\mathbb{Z}_2$ spin liquid with Dirac cones.  We find that including  disorder respecting time-reversal on average and breaking inversion symmetry realizes a phase with degeneracies on, and square-root dispersion near, exceptional rings. 

We consider random magnetic field with zero average.
This type of disorder is different from vacancies~\cite{willans2010, willans2011}, random vortex background~\cite{otten2019}, or nearest neighbor exchange disorder~\cite{Zschocke2015, Knolle2019b} that have been considered in the literature previously.
We treat the random magnetic field perturbatively, assuming that the ground state remains in the zero flux sector.
As explained in the supplemental material (SM)~\ref{app:random_field_model}, this results (at lowest orders) in a Majorana hopping model with disordered first and second neighbor hoppings.
The model preserves time-reversal symmetry on average. To observe exceptional rings, we break inversion symmetry by allowing different magnitude of the random external field on the two sublattices.
This can be achieved experimentally by proximity coupling to a paramagnetic substrate with inequivalent atoms near the two sublattices.

As the symmetry analysis shows, we need to break inversion symmetry. This can be done by introducing disorder potential for next-nearest hopping amplitudes of Majorana modes on orbital $1$, $iV_{\mathbf l}(\mathbf r)\psi_1(\mathbf r+\mathbf l)\psi_1(\mathbf r)$, where the hopping vector can be $\mathbf l=\mathbf a_1,\mathbf a_2,\mathbf a_1+\mathbf a_2$. This term breaks time-reversal symmetry while the exceptional rings require time-reversal symmetry. The strategy is to preserve this symmetry in average. That is to say the statistical average of the time-reversal breaking term vanishes $\langle V_{\mathbf l}(\mathbf r)\rangle=0$. We show in SM~\ref{app:symmetry} 
that the transformation rules in Table~\ref{tb_tsfe} still apply. The disorder is assumed to be uncorrelated at different sites $\langle V_{\mathbf l}(\mathbf r)V_{\mathbf l'}(\mathbf r')\rangle=-F^{\mathbf l}\delta_{\mathbf l,\mathbf l'}\delta(\mathbf r-\mathbf r')$. The self-consistent Born approximation reads as
\begin{equation}
    \Sigma^R_{11}(\omega,\mathbf k)=-\sum_{\mathbf k',\mathbf l} [1-e^{-i(\mathbf k+\mathbf k')\cdot\mathbf l}]F^{\mathbf l}(\mathbf k-\mathbf k')G_{11}(\omega,\mathbf k').
\end{equation}
If $\tilde d_0$ is taken to be infinitesimal, the above equation vanishes as $\omega\to0$. To circumvent this situation, we may consider including a finite $\tilde d_0$ brought by other inversion-symmetry preserving mechanism, such as nearest-neighbour Majorana disorder potentials. This type of disorders brings a small finite lifetime to the Majorana excitations.

In Fig.~\ref{fig:ring} $(c)$ and $(d)$ we show the resulting effective Hamiltonian, with clear evidence of an exceptional ring around the $K$-points.
We also show the single Majorana spectral function in panels $(a)$ and $(b)$, with signatures of a zero-frequency drumhead state inside the exceptional ring.
Somewhat surprisingly, the total density of states, shown in panel $(e)$, has a dip, instead of a peak at zero energy.
This is a consequence of a strong suppression of the zero-frequency spectral weight outside the ER, combined with the larger available phase space outside the ER, allowing the finite-frequency density of states to surpass the zero frequency value.

To support the SCBA results, we use the Kernel Polynomial Method~\cite{Weisse2006, Varjas2019b} to approximate the disorder-averaged retarded Green's function $\langle G^R_{ab} (\omega, {\bf k}) \rangle_{\rm dis}$ numerically.
From this, we obtain the self-energy, the effective Hamiltonian, and the single-particle and two-particle spectral functions, which all show excellent agreement with the SCBA results.  (For details see the SM~\ref{app:numerics}.)
We also show a comparison of the density of states and specific heat~\cite{Feng2020, kao2021} for the clean, nearest-neighbor disordered (without exceptional ring) and second-neighbor disordered (with exceptional ring) systems in Fig.~\ref{fig:ring}~$(e)$ and $(f)$.
The key difference between the two disordered cases is the opposite deviation from the linearized low-temperature specific heat.
This can be understood using Fermi liquid theory and Sommerfeld expansion: the linear term is only sensitive to the zero frequency density of states, however, the sign of the cubic term depends on whether zero frequency is a minimum or a maximum of the density of states. (For details see SM~\ref{app:numerics}.)

\paragraph*{Exceptional points and Fermi arcs in a Kitaev honeycomb model interacting with phonons}

We show here that a $\mathbb{Z}_2$ spin liquid with Dirac cones when coupled to gapped excitations such as optical phonons \cite{metavitsiadis2021optical} instead can realised a qualitatively different phase with point-like exceptional degeneracies connected by Fermi arc degeneracies in the real part of the spectrum. The phonon couplings can be generated by considering vibrations of ions around their equilibrium positions \cite{PhysRevB.101.035103,PhysRevResearch.2.033180}. A spin-spin interaction can be generally written as $\sum_{\mathbf r_1,\alpha,\cdots}J_{\alpha,\beta,\cdots}(\mathbf r_1,\mathbf r_2\cdots)\sigma^\alpha_{\mathbf r_1}\sigma^\beta_{\mathbf r_2}\cdots$. At the equilibrium position $\mathbf r_j=\mathbf r^{(0)}_{j}$, we should have $J_{\alpha,\beta,\cdots}(\mathbf r^{(0)}_1,\mathbf r^{(0)}_2\cdots)=0$ except for those $J$ belonging to the Kitaev interaction. For small ion vibrations, we have the phonon-spin interaction $\sum_{\mathbf r_1,\alpha,j,\cdots}\partial_{\mathbf r_j}J_{\alpha,\beta,\cdots}(\mathbf r_1^{(0)},\mathbf r_2^{(0)}\cdots)\delta\mathbf r_j\sigma^\alpha_{\mathbf r_1}\sigma^\beta_{\mathbf r_2}\cdots$. These spin operators can be written in terms of Majorana operators. The low-energy physics is obtained by those do not excite $\mathbb Z_2$ vortices. For example, a nearest-neighbor Kitaev coupling gives phonons coupled to the different species of Majorana operators $\phi c_1c_2$ and a three-spin coupling can lead to phonons coupled to the same species of Majorana operators $\phi c_1c_1$.

In order to have interesting exceptional degeneracies we need to break inversion symmetry. This can be achieved by giving different couplings to the two sublattices of the honeycomb model. For phonons, this will lead to asymmetric couplings between their normal modes and the two species of Majorana modes. In order to break time-reversal symmetry, we require couplings of the form $\phi c_1c_1$.

In the presence of a Heisenberg interaction and cross interaction, the spin operator gets mixed with the low-energy Majorana operator directly \cite{PhysRevLett.117.037209}. The spin response function can be evident even below the flux gap temperature. We compute the behaviors of the spin structure factor in Fig.~\ref{fig_DPFR}. When the Fermi arc is present, the spin structure constant develops very different directional-dependent shapes compared to the Dirac cones. The cut along the $y-$direction is biased while the cut along the $x-$ direction is uniformly broadened for Fermi arcs. This is very different from the naive isotropic conic structure for Dirac-Majorana dispersion.

\paragraph*{Discussion}
The Kitaev honeycomb model is the subject of much recent interest due to prominent experimental candidates such as $\alpha$-RuCl$_3$~\cite{jackeli2009mott, takagi2019concept, banerjee2017neutron, kitagawa2018spin, kasahara2018majorana, vinkler2018approximately, ye2018quantization, rau2016spin, gohlke2018quantum, hermanns2018}. Recent studies have shown that phonons can play a significant role in the experimental setting~\cite{metavitsiadis2021optical, PhysRevB.101.035103,PhysRevResearch.2.033180}. In this model, the optical phonon energy is well above the flux gap, the signature of exceptional points may be obscured by the effective flux disorders. From our symmetry analysis, we may instead break the inversion symmetry by various means such as stacking the 2D material on a substrate which does not have such a symmetry. The exceptional ring could also be realized by depositing the material on a substrate with random magnetic disorder. Besides measurement of the dynamic structure factor, recent techniques using nonlinear spectroscopy~\cite{PhysRevLett.122.257401, PhysRevLett.124.117205, PhysRevResearch.3.013254, PhysRevResearch.3.L032024, PhysRevX.11.031035} could also provide a more direct probe of the single-particle properties which would more directly reveal the exceptional degeneracies. 

Our results point to new phenomenon in spin liquid depending on the symmetries of the interactions and kinds of disorder present in them. One might have feared that those complications would obfuscate the signature of the spin liquid. Instead, we suggest that it leads to distinctive exceptional degeneracies observable in experimental settings. As our symmetry study works for generic fermionic excitations, the exceptional point and ring can also emerge in other 2D strongly interacting systems. And its generalization to 3D systems may introduce more interesting degeneracies like  knots and links \cite{PhysRevB.99.161115}. 

\acknowledgments{\paragraph{Acknowledgements}- 
KY, DV and EJB acknowledge funding from the Swedish Research Council (VR) and the Knut and Alice Wallenberg Foundation. SM acknowledges funding from the Tsung-Dao Lee Institute. FW is supported in part by the U.S. Department of Energy under grant DE-SC0012567, by the European Research Council under grant 742104, and by the Swedish Research Council under contract 335-2014-7424.
} 

\bibliography{epel}

\newpage
\begin{widetext}

\section{Supplemental Material}
\setcounter{secnumdepth}{2}
\renewcommand{\theequation}{S\arabic{equation}}
\setcounter{equation}{0}
\renewcommand{\thefigure}{S\arabic{figure}}
\setcounter{figure}{0}
\renewcommand{\thetable}{S\arabic{table}}
\setcounter{table}{0}

\subsection{Symmetry transformation of self-energy}
\label{app:symmetry}

In this section we study how the symmetry influences the self-energy. The self-energy is defined as the difference between the interaction/disorder dressed Green's function and the bare Green's function:
\begin{equation}
     \Sigma(\omega)=G_0^{-1}(\omega)-G^{-1}(\omega),
\end{equation}
where $G_0=(\omega-H_0)^{-1}$. In order to find the transformation of the self-energy, we compute the transformation of the Green's function. The retarded and advanced Green's functions are defined as
\begin{equation}
    G^R_{ab}(t,\mathbf k)=-i\theta(t)\langle \psi_{a,\mathbf k}(t)\psi^\dagger_{b,\mathbf k}(0)\rangle+\langle\psi^\dagger_{b,\mathbf k}(0)\psi_{a,\mathbf k}(t)\rangle,\qquad  G^A_{ab}(t,\mathbf k)=i\theta(-t)\langle \psi_{a,\mathbf k}(t)\psi^\dagger_{b,\mathbf k}(0)\rangle+\langle\psi^\dagger_{b,\mathbf k}(0)\psi_{a,\mathbf k}(t)\rangle.\label{eq_arg}
\end{equation}
Here we use the general notion of complex fermions. For Majoranas, we can simply replace $\psi_{a,\mathbf k}=c_{a,\mathbf k}$ and $\psi^\dagger_{a,\mathbf k}=c_{a,-\mathbf k}$. After doing the Fourier transformation, we obtain their spectral decomposition as
\begin{align}
    G^R_{ab}(\omega,\mathbf k)&=\sum_{mn}\frac{X_{ab}^{mn}(\mathbf k)}{\omega-(E_m-E_n)+i0^+}\frac{e^{-E_n/T}}{Z}+\sum_{mn}\frac{X_{ab}^{nm}(\mathbf k)}{\omega+(E_m-E_n)+i0^+}\frac{e^{-E_n/T}}{Z},\qquad \\
    G^A_{ab}(\omega,\mathbf k)&=\sum_{mn}\frac{X_{ab}^{mn}(\mathbf k)}{\omega-(E_m-E_n)-i0^+}\frac{e^{-E_n/T}}{Z}+\sum_{mn}\frac{X_{ab}^{nm}(\mathbf k)}{\omega+(E_m-E_n)-i0^+}\frac{e^{-E_n/T}}{Z}.
\end{align}
The $i0^+$ term is included to insure the correct causality: $G^{R/A}(\omega)$ has to be analytic for $\textrm{Im}\omega>0$ or $\textrm{Im}\omega<0$ respectively. The weight $X_{ab}^{mn}$ is given by $\langle n|\psi_{a,\mathbf k}|m\rangle\langle m|\psi^\dagger_{b,\mathbf k}|n\rangle$ and $Z$ is the partition function. From this we observe that $G^{R/A}_{ab}$, regarded as a matrix with indices $ab$, can be expressed as an Hermitian matrix $\bar G_{ab}$ plus/minus an anti-Hermtian matrix $i\tilde G_{ab}$:
\begin{equation}
    G^{R/A}_{ab}(\omega,\mathbf k)=\bar G_{ab}(\omega,\mathbf k)\pm i\tilde G_{ab}(\omega,\mathbf k).
\end{equation}
Furthermore, $G^{R}(\omega)$ with complex $\omega$ can be related to $G^{A}(\omega)$ as follows:
\begin{equation}
    G^{R}_{ab}(\omega,\mathbf k)=[G^{A}_{ba}(\omega^\ast,\mathbf k)]^\ast.
\end{equation}
According to the definition the self-energy, $\Sigma^R(\omega)$ and $\Sigma^A(\omega)$ satisfy similar rules and their Hermitian and anti-Hermitian components are defined accordingly.

We then look at symmetry transformations. For inversion symmetry and other spatial symmetries, the transformation of the self-energy exactly follows the transformation of the Hamiltonian. Assume under inversion symmetry, $\mathcal P\psi_{a,\mathbf k}\mathcal P^{-1}=\sum_b U^I_{ab}\psi_{b,-\mathbf k}$. We can easily deduce $G^R_{ab}(\omega,\mathbf k)=\sum_{cd}U^I_{ac}U^{I\ast}_{bd}G^R_{cd}(\omega,-\mathbf k)$. Particle-hole symmetry interchanges the creation and annihilation operators
\begin{equation}
     G^{R}_{ab}(t,\mathbf k)=-G^{A}_{ba}(-t,-\mathbf k)\Rightarrow  G^{R}_{ab}(\omega,\mathbf k)=-G^{A}_{ba}(-\omega, -\mathbf k).
\end{equation}
Time-reversal symmetry acts differently from other symmetries as there can be Krammer degeneracy where $\mathcal T|n\rangle\ne |n\rangle$. Nevertheless the Krammer pairs are usually summed with equal weight in equilibrium the density matrix. The density matrix is still invariant under time-reversal symmetry $\sum_n |\mathcal Tn\rangle\langle \mathcal T n| f(E_n)=\sum_n |n\rangle\langle n| f(E_n)$, where $f(E_n)$ is the weight function corresponding to the density matrix. So we find 
\begin{align}
    \sum_n f(n)\langle n|\psi_{a,\mathbf k}(t)\psi^\dagger_{b,\mathbf k}(0)|n\rangle&=\sum_n f(n)\langle n|\mathcal T^{-1}\mathcal T\psi_{a,\mathbf k}(t)\mathcal T^{-1}\mathcal T\psi^\dagger_{b,\mathbf k}(0)\mathcal T^{-1}\mathcal T|n\rangle\nonumber \\
    &=\sum_{n, cd} f(n)U_{ac}U^\ast_{bd} \langle n|\mathcal T^{-1}\psi_{c,-\mathbf k}(-t)\psi^\dagger_{d,-\mathbf k}(0)|\mathcal Tn\rangle\nonumber\\
    &=\sum_{n,cd} f(n)U_{ac}U^\ast_{bd} \langle \mathcal Tn|\mathcal \psi_{c,-\mathbf k}(-t)\psi^\dagger_{d,-\mathbf k}(0)|\mathcal Tn\rangle^\ast\nonumber\\
    &=\sum_{cd,n}U_{ac}U^\ast_{bd}  f(n)\langle n|\mathcal \psi_{c,-\mathbf k}(-t)\psi^\dagger_{d,-\mathbf k}(0)|n\rangle^\ast\nonumber.
\end{align}
The third line comes from the property of an anti-unitary operator. With the help of this equation, we obtain the rule under time-reversal transformation
\begin{equation}
    G^{R}_{ab}(t,\mathbf k)=\sum_{cd} U_{ac}U^\ast_{bd}[G^{A}_{cd}(-t,-\mathbf k)]^\ast\Rightarrow  G^{R}_{ab}(\omega,\mathbf k)=\sum_{cd} U_{ac}U^\ast_{bd}[G^{A}_{cd}(\omega^\ast,-\mathbf k)]^\ast,
\end{equation}
where in the second expression we analytically continue $G^{R/A}(\omega)$ to $\textrm{Im}\omega>0$ and $\textrm{Im}\omega<0$ respectively. Since $U,U^I$ are unitary matrices, the transformation rules of the self-energy immediately follow.

The above calculation shows how the interacting Green's function transforms under symmetries. We can also prove similarly the transformation rules for the disorder-averaged Green's function. To include disorder averaging, we denote the eigenstates and eigenvalues of the disordered Hamiltonian $H(V)$ as $|n(V)\rangle$ and $E(V)$. In the density matrix, each eigenstate needs to be summed with the disorder distribution weight $g(V)$. We call that a symmetry is preserved in average if the Hamiltonian transforms under the symmetry $\mathcal S$ as $\mathcal S H(V)\mathcal S^{-1}=H(SV)$ and $g(V)=g(SV)$. Using the equation $\mathcal S H(V)\mathcal S^{-1}\mathcal S|n(V)\rangle=E_n(V)\mathcal S|n(V)\rangle$, we find that $\mathcal S|n(V)\rangle$ is an eigenstate of $H(SV)$ with eigenvalue $E_n(V)$. The  eigenstates can thus be denoted as $E_n(SV)=E_n(V)$ and $\mathcal S|n(V)\rangle=|n(SV)\rangle$. The partition functions related by $V$ are also equal and the temperature averaging is denote as $f[E_n(V),V]=f[E_n(SV),SV]$. For simplicity we take $\mathcal S$ to be unitary and $\mathcal S\psi_a\mathcal S^{-1}=\sum_b U^S_{ab}\psi_b$. The expectation values behave as
\begin{align}
    &\sum_{n,V} g(V)f[E_n(V),V]\langle n(V)|\psi_{a}(t,V)\psi^\dagger_{b}(0)|n(V)\rangle\nonumber\\
    =&\sum_{n,V} g(V)f[E_n(V),V]\langle n(V)|\mathcal S^{-1}\mathcal S\psi_a(t,V)\mathcal S^{-1}\mathcal S\psi^\dagger_b(0)\mathcal S^{-1}\mathcal S|n(V)\rangle\nonumber\\
    =&\sum_{n,V,cd} g(V)f[E_n(V),V]U^{S}_{ac}U^{S\ast}_{bd}\langle n(SV)|\psi_c(t,SV)\mathcal \psi^\dagger_d(0)\mathcal |n(SV)\rangle\nonumber\\
    =&\sum_{n,V,cd} g(SV)f[E_n(SV),SV]U^{S}_{ac}U^{S\ast}_{bd}\langle n(SV)|\psi_c(t,SV)\mathcal \psi^\dagger_d(0)\mathcal |n(SV)\rangle\nonumber\\
    =&\sum_{n,V,cd} g(V)f[E_n(V),V]U^{S}_{ac}U^{S\ast}_{bd}\langle n(V)|\psi_c(t,V)\mathcal \psi^\dagger_d(0)\mathcal |n(V),
\end{align}
where in the last equation we change the sum over $V$ to $S^{-1}V$. This proves that the disorder-averaged Green's function transforms in the same way as the interacting Green's function as long as the (unitary) symmetry is preserved in average. For anti-unitary symmetries, the proof is similar and we do not list out the details.

\subsection{General rules for exceptional degeneracy and Dirac fermions}

As we have shown in our previous work \cite{PhysRevLett.126.077201}, exceptional degeneracy is general and not limited to Dirac parent Hamiltonians. The $2\times 2$ Hamiltonian, as in the main text, can always be parameterized by the Pauli matrix with complex vectors $\{d_0(\mathbf k),\mathbf d(\mathbf k)\}$. The eigenvalues are given by $d_0(\mathbf k)\pm \sqrt{\mathbf d(\mathbf k)\cdot\mathbf d(\mathbf k)}$. The system is gapped as long as $|\mathbf d(\mathbf k)\cdot\mathbf d(\mathbf k)|\ne 0$ while gapless when $|\mathbf d(\mathbf k)\cdot\mathbf d(\mathbf k)|= 0$. Note $|\mathbf d(\mathbf k)\cdot\mathbf d(\mathbf k)|\ne |\mathbf d(\mathbf k)|^2$ when $\mathbf d(\mathbf k)$ is a complex vector.

Now we start with a gapped Hermitian system where $\mathbf d^{(0)}(\mathbf k)$ are real and there is a minimal value $\Delta=|\mathbf d^{(0)}(\mathbf k_\ast)\cdot\mathbf d^{(0)}(\mathbf k_\ast)|$ at $\mathbf k_\ast$. Without loss of generality, we take $\mathbf d^{(0)}(\mathbf k_\ast)=\{0,0,\Delta \}$. This is the situation when a magnetic field $\mathbf h\cdot\boldsymbol{\sigma}$ along the $(1,1,1)$ direction is applied to the Kitaev model \cite{KITAEVHoneycomb}. In order to close the gap, we may choose a complex $\mathbf d^{(1)}(\mathbf k)=\{i \Delta,0,0\}$, which may come from the selfenergy. Then the new non-Hermitian Hamiltonian $H=d_0I+(\mathbf d^{(0)}+\mathbf d^{(1)})\cdot\boldsymbol\tau$ is gapless at $\mathbf k_\ast$. The Hermitian gap $\Delta$ is closed by the non-Hermitian band attraction effect. Moreover, this effect does not need fine tuning. If we perturb $\mathbf d^{(1)}$ slightly while keeping its absolute value larger than $\Delta$, we can always find in the neighborhood of $\mathbf k_\ast$ a series of $\mathbf k'$ where $\mathbf d^{(1)}\cdot \mathbf d^{(0)}(\mathbf k')=0$. This only needs $\mathbf k'$ to lie on some line through $\mathbf k_\ast$. By fixing the position of $\mathbf k'$ along this line, we can further require  $|\mathbf d^{(0)}(\mathbf k')|=|\mathbf d^{(1)}|$. This shows that exceptional degeneracy is a general phenomenon and we can obtain it from either a gapped or a gapless system.

We may also consider Dirac fermions satisfying the same symmetry transformation rules. The results naturally follow from the previous section. As complex fermions now do not have particle-hole symmetry, we can either break the inversion symmetry or the particle-hole symmetry to obtain exceptional points with non-trivial non-Hermitian selfenergy. The particle-hole symmetric case is the same as discussed in the main text. If particle-hole symmetry is broken while inversion symmetry conserved, the exceptional ring is no longer protected by time reversal symmetry, and we generically get exceptional points. Time-reversal symmetry and inversion symmetry forces the component $d_z$ to be zero. As $d_x$ and $d_y$ also have Hermitian components, we still have exceptional points when  time-reversal symmetry and inversion symmetry are conserved. To conclude, with inversion symmetry, we often obtain exceptional points.

\subsection{Details of the interaction induced self-energy}

We assume that the Majorana is coupled to a real bosonic field $\phi$ that conveys the interaction. In physical systems, examples of $\phi$ can be additional degrees of freedom such as phonons of the ions, or collective modes in the substrate like magnons. The most general leading order term should be
\begin{equation}
    H_{bm}=i\sum_{\mathbf r,\mathbf l_1,\mathbf l_2,a,b}\psi_{a,\mathbf r+\mathbf l_1}\lambda^{ab}_{\mathbf  l_1,\mathbf l_2} \psi_{b,\mathbf r+\mathbf l_2} \phi_{\mathbf r},
\end{equation}
with $\lambda^{ab}_{\mathbf  l_1,\mathbf l_2}$ being real coupling constants and anti-symmetric under the exchange $(a,\mathbf l_1)\leftrightarrow (b,\mathbf l_2)$. In momentum space this interaction vertex is expressed as
\begin{equation}
    H_{bm}=i\sum_{\mathbf k,\mathbf q,a,b}\psi^\dagger_{a,\mathbf k+\mathbf q}\lambda^{ab}_{\mathbf k+\mathbf q,\mathbf k} \psi_{b,\mathbf k} \phi_{\mathbf q},\quad \lambda^{ab}_{\mathbf k+\mathbf q,\mathbf k}=\sum_{\mathbf l_1,\mathbf l_2}\lambda^{ab}_{\mathbf  l_1,\mathbf l_2} e^{-i(\mathbf q+\mathbf k)\cdot\mathbf l_1+i\mathbf k\cdot\mathbf l_2},\label{eq_para_bm}
\end{equation}
where $\psi^\dagger_{\mathbf k}=\psi_{-\mathbf k}$ and $\phi^\dagger_{\mathbf q}=\phi_{-\mathbf q}$ as Majorana and the bosonic field are assumed to be real variables. The PH symmetry of Majoranas requries $\lambda^{ab}_{\mathbf q+\mathbf k,\mathbf k}=-\lambda^{ba}_{-\mathbf k,-\mathbf q-\mathbf k}$ and Hermiticity requires $\lambda^{ab}_{\mathbf q+\mathbf k,\mathbf k}=(\lambda^{ab}_{-\mathbf q-\mathbf k,-\mathbf k})^\ast$. So $\lambda^{ab}_{\mathbf q+\mathbf k,\mathbf k}=-(\lambda^{ba}_{\mathbf k,\mathbf q+\mathbf k})^\ast$. The $\phi$ here is taken as one of the eigenmode of the bosonic field, as the bosonic field can have several degrees of freedom inside a unit cell. The matrix $\lambda^{ab}$ is off-diagonal if the bosons only couple to nearest-neighbour spins and can have diagonal components when multiple spins are coupled. The propagators of the Majoranas and the bosonic field at zero temperature can be parametrized as
\begin{align}
    G_{ab}(\omega,\mathbf k)&=-i\int \langle \mathcal T\psi_{a,\mathbf k}(t)\psi^\dagger_{b,\mathbf k}(0)\rangle e^{i\omega t}=\sum_{s=\pm 1} \frac{B_{ab,s}(\mathbf k)}{\omega+s|\varepsilon_{\mathbf k}|-si 0^+},\\
    D(\Omega,\mathbf q)&=-i\int \langle \mathcal T \phi_{\mathbf q}\phi_{-\mathbf q}\rangle e^{i\omega t}=\frac{C}{\Omega^2-\Omega_{\mathbf q}^2+i0^\dagger}=\sum_s \frac{sC}{2\Omega_{\mathbf q}(\Omega-s\Omega_{\mathbf q}+si0^\dagger)},
\end{align}
where $2B_s=I-s\mathbf d\cdot\boldsymbol\sigma/\varepsilon_{\mathbf k}$ and $\varepsilon_{\mathbf k}$ is the aboslute value of the unperturbed Majorana energy. The constant $C$ relies on the detailed dispersion of the bosonic field. The symbol $\Omega_{\mathbf q}$ denotes the absolute value of the boson energy. In order to carry out the finite-temperature calculation, we need to rotate the real frequencies in the above equations to imaginary Matsubara frequencies, $\omega\to i\omega_n=(2n+1)\pi/T, \Omega\to i\Omega_n=2n \pi/T$ with $T$ the temperature. The boson contribution to the self-energy can be written as:
\begin{align}
    \Sigma_{ab}(i\omega_n,\mathbf k)&=\frac{1}{\beta}\sum_{\Omega_m,\mathbf q,c,d}\lambda^{ac}_{\mathbf k,\mathbf k+\mathbf q}G_{cd}(i\omega_n+i\Omega_m,\mathbf k+\mathbf q)\lambda^{db}_{\mathbf k+\mathbf q,\mathbf k} D(i\Omega_m,\mathbf q)\\
    &=\frac{1}{\beta}\sum_{\Omega_m,\mathbf q,s,s',c,d}\frac{1}{2\Omega_{\mathbf q}}\lambda^{ac}_{\mathbf k,\mathbf k+\mathbf q}\frac{B_{cd,s}(\mathbf k+\mathbf q)}{i\omega_n+i\Omega_m+s|\varepsilon_{\mathbf k+\mathbf q}|}\lambda^{db}_{\mathbf k+\mathbf q,\mathbf k}  \frac{s'C}{i\Omega_m-s'\Omega_{\mathbf q}}\\
    &=-\sum_{\mathbf q,s,s',c,d}\frac{1}{2\Omega_{\mathbf q}}\lambda^{ac}_{\mathbf k,\mathbf k+\mathbf q}B_{cd,s}(\mathbf k+\mathbf q)\lambda^{db}_{\mathbf k+\mathbf q,\mathbf k}\frac{n_B(s'\Omega_{\mathbf q})+n_F(-s|\varepsilon_{\mathbf k+\mathbf q}|)}{s'\Omega_{\mathbf q}+i\omega_n+s|\varepsilon_{\mathbf k+\mathbf q}|} s'C,
\end{align}
where $n_B(z)=[\coth(z\beta/2)-1]/2,n_F(z)=[1-\tanh(z\beta/2)]/2$ are the Bose and Fermi distributions and $\beta=1/T$. To go back to real frequency, we plug in $i\omega_n\to \omega+i0^+$. The NH part is only non-vanishing when the denominator is zero:
\begin{align}
    \frac{i\tilde\Sigma^R_{ab}(\omega,\mathbf k)}{\pi}=&\sum_{\mathbf q,s,s',c,d}\frac{s'C}{2\Omega_{\mathbf q}}\lambda^{ac}_{\mathbf k,\mathbf k+\mathbf q}B_{cd,s}(\mathbf k+\mathbf q)\lambda^{db}_{\mathbf k+\mathbf q,\mathbf k}\left[n_B(s'\Omega_{\mathbf q})+n_F(-s|\varepsilon_{\mathbf k+\mathbf q}|)\right]\delta(s'\Omega_{\mathbf q}+\omega+s|\varepsilon_{\mathbf k+\mathbf q}|)\\
    =&\sum_{\mathbf q,s,s',c,d}\frac{s'C}{4\Omega_{\mathbf q}}\lambda^{ac}_{\mathbf k,\mathbf k+\mathbf q}B_{cd,s}(\mathbf k+\mathbf q)\lambda^{db}_{\mathbf k+\mathbf q,\mathbf k}\left[s'\coth\left(\frac{\Omega_{\mathbf q}\beta}{2}\right)+s\tanh\left(\frac{|\varepsilon_{\mathbf k+\mathbf q}|\beta}{2}\right)\right]\\
    &\times\delta(s'\Omega_{\mathbf q}+\omega+s| \varepsilon_{\mathbf k+\mathbf q}|).
\end{align}
Let $\omega\to0$ and focus on the self-energy around the gapless point $\varepsilon_{\mathbf k_\ast}= 0$. The Majorana energy is described by $\varepsilon_{\mathbf k_\ast+\mathbf q}=v_Fq$. The Dirac-$\delta$ function is non-vanishing only for $s=-s'$. Notice that if the boson is gapless at $\mathbf q=0$, the phase space for decaying channels vanishes at $\omega=0$. The above function either vanishes or can have singular behaviours near zero frequency around the Dirac point $\mathbf k_\ast$ \cite{PhysRevB.73.220503}. To avoid this, we may consider gapped bonsons, such as optical phnons, whose energy can be approximated by a constant $\Omega_{\mathbf q}=\Delta_b$ for small $\mathbf q$. The integral along the radius of $q$ can be carried easily and the NH components of the self-energy are approximated in the Dirac-cone limit by
\begin{align}
   i\tilde\Sigma^R_{ab}(0,\mathbf k_\ast)=&\sum_{ s,c,d}\int d\theta \frac{C}{16v^2_F\pi}\lambda^{ac}_{\mathbf k_\ast,\mathbf q_\theta}B_{cd,s}(\mathbf q_\theta)\lambda^{db}_{\mathbf q_\theta,\mathbf k_\ast} \left[\coth\left(\frac{\Delta_b\beta}{2}\right)-\tanh\left(\frac{\Delta_b\beta}{2}\right)\right]=\frac{C'_{ab}}{\exp(\Delta_b\beta)-\exp(-\Delta_b\beta)}\\
   C'_{ab}=&\sum_{c}\int d\theta \frac{C}{4v^2_F\pi}\lambda^{ac}_{\mathbf k_\ast,\mathbf q_\theta}\lambda^{cb}_{\mathbf q_\theta,\mathbf k_\ast}=-\sum_{c}\int d\theta \frac{C}{4v^2_F\pi}\lambda^{ac}_{\mathbf k_\ast,\mathbf q_\theta}(\lambda^{bc}_{\mathbf k_\ast,\mathbf q_\theta})^\ast,
\end{align}
where $\mathbf q_\theta=\mathbf k_\ast+\hat{\mathbf e}_\theta\Delta_b/v_F$ with $\hat{\mathbf e}$ the unit vector of angle $\theta$ with respect to the $x$-axis and we make use of the relation $\sum_s B_{cd,s}=\delta_{cd}$ and Hermiticity of $\lambda^{ab}_{\mathbf k,\mathbf q}$. 

We need to find out the phonon-spin interaction vertex $\lambda^{ab}_{\mathbf k,\mathbf q}$ that can generate non-trivial NH self-energy. A spin-spin interaction can be generally written as $\sum_{\mathbf r_1,\alpha,\cdots}J_{\alpha,\beta,\cdots}(\mathbf r_1,\mathbf r_2\cdots)\sigma^\alpha_{\mathbf r_1}\sigma^\beta_{\mathbf r_2}\cdots$. At the equilibrium position $\mathbf r_j=\mathbf r^{(0)}_{j}$, we should have $J_{\alpha,\beta,\cdots}(\mathbf r^{(0)}_1,\mathbf r^{(0)}_2\cdots)=0$ except for those $J$ belonging to the Kitaev interaction. The bosons are coupled to the spin via $\sum_{\mathbf r,\mathbf r_1,\alpha,\cdots}\Lambda_{\alpha,\beta,\cdots}(\mathbf r,\mathbf r_1,\mathbf r_2\cdots)\phi(\mathbf r)\sigma^\alpha_{\mathbf r_1}\sigma^\beta_{\mathbf r_2}\cdots$. At low energy, the most relevant spin-spin interactions are those not creating a flux excitations. The Majorana-boson vertex is obtained by project these expressions to the zero-flux sector. The simplest one is to couple the boson to the nearest-neighbour Kitaev interaction. This gives in the Majorana representation an off-diagonal component $\lambda^{12}_{\mathbf k_\ast,\mathbf q_\theta}=-(\lambda^{21}_{\mathbf q_\theta,\mathbf k_\ast})^\ast$. The NH self-energy thus has components $\tilde\Sigma_{11}\sim \int d\theta|\lambda^{12}_{\mathbf k_\ast,\mathbf q_\theta}|^2$ and $\tilde\Sigma_{22}\sim \int d\theta|\lambda^{21}_{\mathbf k_\ast,\mathbf q_\theta}|^2$. When we have inversion symmetry, $\lambda^{ab}_{\mathbf k+\mathbf q,\mathbf k}=\lambda^{ba}_{-\mathbf k-\mathbf q,-\mathbf k}$, the terms $\tilde\Sigma_{11}$ and $\tilde\Sigma_{22}$ are equal as expected from the symmetry transformation of the self-energy. We can also prove this explicitly using the transformation $|\lambda^{21}_{\mathbf k_\ast,\mathbf q_\theta}|=|\lambda^{12}_{-\mathbf k_\ast,-\mathbf q_\theta}|=|\lambda^{12}_{\mathbf k_\ast,\mathbf q_\theta}|$. To break inversion symmetry, we may consider the situation when the bosons couples differently to the two orbitals in the honeycomb lattice. As inversion transformation needs to swap the two orbitals, such a coupling would break inversion symmetry. In this situation, we can have different $\tilde\Sigma_{11}$ and $\tilde\Sigma_{22}$ and therefore obtain an exceptional ring. Using the notation in Eq.~\eqref{eq_para_bm}, we can express their difference as
\begin{equation}
    \tilde d_z(\mathbf k_\ast)=\frac{\tilde\Sigma_{11}(0,\mathbf k_\ast)-\tilde\Sigma_{22}(0,\mathbf k_\ast)}{2}=\sum_{\mathbf l_1,\mathbf l_2,\mathbf l'_1,\mathbf l'_2}\frac{C\left(\lambda^{12}_{\mathbf l_1,\mathbf l_2}\lambda^{12}_{\mathbf l'_1,\mathbf l'_2}-\lambda^{12}_{\mathbf l_2,\mathbf l_1}\lambda^{12}_{\mathbf l'_2,\mathbf l'_1}\right)}{16v^2_F\sinh\left(\Delta_b\beta\right)}\cos[\mathbf k_\ast\cdot(\mathbf l_1-\mathbf l_2+\mathbf l'_2-\mathbf l'_1)]J_0\left(\frac{\Delta_b}{v_F}|\mathbf l_2-\mathbf l'_2|\right),
\end{equation}
where $J_0$ is the zeroth Bessel's function.

In order to obtain exceptional points, we need to further consider multiple-spin interactions. A three-spin interaction $\sigma^x_j\sigma^y_k\sigma^z_l$ translates to $\psi_{1,j}\psi_{1,k}$ and $\psi_{2,j}\psi_{2,k}$ in the Majorana representation. This may be obtained by considering what terms a time-reversal symmetry interaction can generate in the zero-flux sector. These terms contribute diagonal components  $\lambda^{11}_{\mathbf k_\ast,\mathbf q_\theta}$  and $\lambda^{22}_{\mathbf k,\mathbf q}$ to the phonon-Majorana couplings. To break inversion symmetry, we may set only $\lambda^{11}_{\mathbf k_\ast,\mathbf q_\theta}$ non-zero. When they are mixed with the nearest-neighbour contributions, we can have $i\tilde\Sigma_{12}\sim\int d\theta(\lambda^{11}_{\mathbf k_\ast,\mathbf q_\theta}\lambda^{12}_{\mathbf q_\theta,\mathbf k_\ast})$ and $i\tilde\Sigma_{21}\sim \int d\theta(\lambda^{21}_{\mathbf k_\ast,\mathbf q_\theta}\lambda^{11}_{\mathbf q_\theta,\mathbf k_\ast})$. This form of self-energy would admit exceptional points. However, notice that time-reversal symmetry has now been broken. A gap can be opened with the same order of magnitude $\bar\Sigma^{11}\sim |\lambda^{11}\lambda^{21}|$. There is a competition between the band attraction due to the non-trivial NH self-energy and gap opening coming from the Hermitian self-energy. We need to choose the parameters carefully so that the NH components prevail in order to observe the exceptional points. In practice, we may mix several time-reversal breaking terms together so that their sum for $\bar d_z$ becomes minimal compared to the non-trivial NH components.

An example of gapped bosons is the optical phonon. A possible way to break inversion symmetry is to give the ions at the two orbitals different masses, which is actually a very natural assumption for the existence of optical phonons. Then the normal mode of the optical phonon couples different to the spins at the two in-equivalent orbitals. This breaks inversion symmetry and is possible to generate exceptional rings or exceptional points.

\subsection{Self-consistent Born calculation}

We use the self-consistent Born approximation to evaluate the self-energy brought by uncorrelated next-to-nearest neighbor (NNN) hopping terms. The disorder is parameterized by
\begin{equation}
    \hat V=\sum_{\mathbf r,\mathbf l,a,b} iV_{ab,\mathbf l}(\mathbf r) \psi_a(\mathbf r+\mathbf l)\psi_b(\mathbf r).
\end{equation}
We explicit mark that the disorder is along a link of direction $\mathbf l$ from Majorana $b$ to Majorana $a$ and particle-hole symmetry tells us $V_{ab,\mathbf l}(\mathbf r)=-V_{ba,-\mathbf l}(\mathbf r+\mathbf l)$. We assume translationally invariant disorders and the disorder correlation function is given by
\begin{align}
    \langle  V_{ab,\mathbf l}(\mathbf r)  V_{cd,\mathbf l'}(\mathbf r')\rangle&=-\langle  V_{ab,\mathbf l}(\mathbf r)  V_{dc,-\mathbf l'}(\mathbf r'+\mathbf l')\rangle=-\langle  V_{ba,-\mathbf l}(\mathbf r+\mathbf l)  V_{cd,\mathbf l'}(\mathbf r')\rangle=\langle  V_{ba,-\mathbf l}(\mathbf r+\mathbf l) V_{dc,-\mathbf l'}(\mathbf r'+\mathbf l')\rangle\nonumber\\
    &=\delta_{\mathbf l+\mathbf l'}F^{\mathbf l}_{abcd}(\mathbf r-\mathbf r'+\mathbf l)-\delta_{\mathbf l-\mathbf l'}F^{\mathbf l}_{abdc}(\mathbf r-\mathbf r')=\delta_{\mathbf l+\mathbf l'} F^{-\mathbf l}_{badc}(\mathbf r-\mathbf r'+\mathbf l)-\delta_{\mathbf l-\mathbf l'}F^{-\mathbf l}_{bacd}(\mathbf r-\mathbf r').
\end{align}
In the second line, the correlation symbol $F^{\mathbf l}_{abcd}$ has the property $F^{\mathbf l}_{abcd}(\mathbf r)=F^{-\mathbf l}_{badc}(\mathbf r)$. The two expressions in the second line correspond to two ways of marking the disorder: we can choose the direction of hopping to be along either $\mathbf l$ or $-\mathbf l$. The uncorrelated disorder restricts that only the disorder potentials acting on the same link $(a,\mathbf r+\mathbf l)-(b,\mathbf r)$ have non-vanishing dipole moment, $F^{\mathbf l}_{abcd}(\mathbf r)=F^{\mathbf l}\delta_{ad}\delta_{bc}\delta(\mathbf r)$ with $F^{\mathbf l}\le 0$. In coordinate space, we have
\begin{equation}
    \Sigma_{ab}(\omega,\mathbf r,\mathbf r')=\sum_{c,\mathbf l,d,\mathbf l'}\langle V_{ac,\mathbf l}(\mathbf r-\mathbf l)V_{db,\mathbf l'}(\mathbf r')\rangle G_{cd}(\omega,\mathbf r-\mathbf l,\mathbf r'+\mathbf l').
\end{equation}
In the above equation, we sum each bond disorder twice, i.e. including both $\pm\mathbf l$ terms. This accounts for the fact that both the two Majorana operators in $iV_{ab,\mathbf l}(\mathbf r) \psi_a(\mathbf r+\mathbf l)\psi_b(\mathbf r)$ contribute to the equation of motion, unlike complex fermions. Using the correlation function of the disorder, the Majorana self-energy satisfies the self-consistent equation
\begin{align}
    \Sigma_{ab}(\omega,\mathbf k)=-\sum_{cd,\mathbf k',\mathbf l} F^{\mathbf l}_{acdb}(\mathbf k-\mathbf k')G_{cd}(\omega,\mathbf k')+\sum_{cd,\mathbf k',\mathbf l}e^{-i(\mathbf k+\mathbf k')\cdot\mathbf l} F^{\mathbf l}_{acbd}(\mathbf k-\mathbf k')
     G_{cd}(\omega,\mathbf k'),\label{eq_scba}
\end{align}
As before, we need to sum both directions of a link $\mathbf l$ and $-\mathbf l$ in the above equation. The first summation in the above equation is the usual convolution result for a fermionic system. It comes from the disorder correlation with links along reciprocal directions. In the second sum, the links in the disorder correlation function take the same direction and there is an explicit $\mathbf k$ dependence even if the disorder is uncorrelated. 

The above matrix equation can be expressed as a multi-variable integral equation in terms of  $d_0,\mathbf d$. A numerical solution can be obtained by iteration. For NNN uncorrelated disorders, the integral equation takes a simple form and is solvable under the Dirac-cone approximation. In this situation, the disorder correlation function is a constant in momentum space has only the $(1111)$ component $F^{\mathbf l}_{1111}(\mathbf k)=F\delta_{\mathbf l,\pm\mathbf a_1}$, where $\mathbf a_1$ is a lattice vector. The momentum scale we consider is much smaller than the reciprocal lattice so that we can further approximate the phase with their values at the Dirac point $\exp[-i(\mathbf k+\mathbf k')\cdot\mathbf l]\simeq \exp(\pm i2\mathbf k_\ast\cdot\mathbf a_1)$ in Eq.~\eqref{eq_scba}, where the Dirac point position is $\mathbf k_\ast=(2\pi/3,2\pi/\sqrt{3})/a$. Linearizing the Hamiltonian near the Dirac point by redefining $\mathbf k-\mathbf k_\ast\to\mathbf k$, we parameterize the self-energy and the Green's function as
\begin{equation}
    H_0(\mathbf k)+\Sigma^R(\omega,\mathbf k)=\left(\begin{array}{cc}
        \Sigma_{1}(\omega)-i\tau & \mathbf v_x\cdot\mathbf k-i\mathbf v_y\cdot\mathbf k \\
        \mathbf v_x\cdot\mathbf k+i\mathbf v_y\cdot\mathbf k &  -i\tau
    \end{array}\right),\
     G^R_{11}=\frac{\omega+i\tau}{[\omega-\Sigma_{1}(\omega)+i\tau](\omega+i\tau)-[(\mathbf v_x\cdot\mathbf k)^2+(\mathbf v_y\cdot\mathbf k)^2]},
\end{equation}
where the velocity are given by $\mathbf v_{x/y}=\partial_{\mathbf k}d_{x/y}(\mathbf k)$ at the Dirac point and $\tau$ is a lifetime represented by an imaginary identity operator. At the isotropic point $J_x=J_y=J_z$, we can put $\mathbf v_x\perp\mathbf v_y$ and $|\mathbf v_x|=|\mathbf v_y|=v$. Plugging the above expression into the self-consistent equation, we obtain
\begin{equation}
    \Sigma_{1}(\omega)=-\sum_{\mathbf k'}\frac{3(\omega+i\tau)F}{[\omega-\Sigma_{1}(\omega)+i\tau](\omega+i\tau)-v^2k'^2}
\end{equation}
After integrating out the angular part, the expression becomes
\begin{equation}
    \Sigma_{1}(\omega)=\frac{3(\omega+i\tau) F }{4\pi v^2}\ln\left(\frac{\Lambda}{Z}\right)
    ,\quad Z=\omega^2-[\Sigma_{1}(\omega)-2i\tau]\omega-i\tau[\Sigma_{1}(\omega)-i\tau],\label{eq_snDSCBA}
\end{equation}
where $\Lambda$ is the cutoff of the Dirac-cone approximation (unit $\omega^2$). Notice that in the continuum Dirac model, $F$ now takes the unit $\omega^2/k^2$. This is a transcendental equation for $\Sigma_1(\omega)$. The logarithm has to be taken with caution as the integral is in general performed on the complex plane. At $\omega=0$, we can take $\Sigma_{1}(0)$ to be purely imaginary and solve it numerically. For small $\tau$, we expect $\Sigma_{1}(0)$ to be proportional to the square of the disorder strength. We verify that it admits a physical solution for various parameters chosen in Fig~\ref{fig_SCBAD}. The ratio between the imaginary parts of $d_z$ and $d_0$ is also obtained.

\begin{figure}[t]
\textbf{\textsf{a}} \raisebox{-0.9\height}{\includegraphics[width=0.45\linewidth]{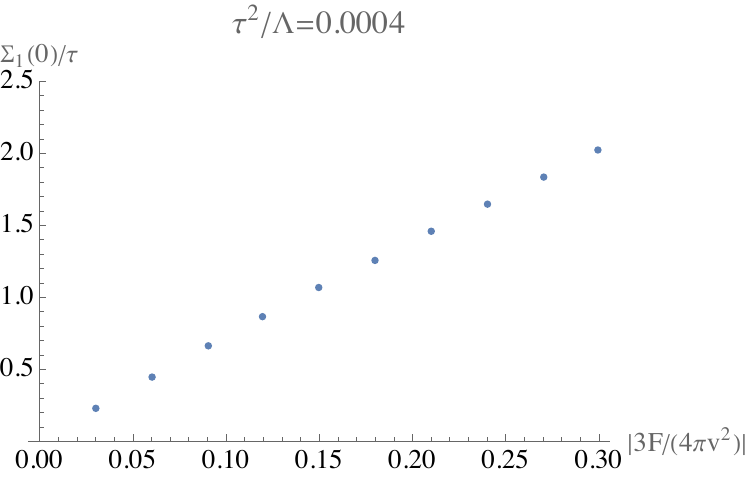}}
\
\textbf{\textsf{b}} \raisebox{-0.9\height}{\includegraphics[width=0.45\linewidth]{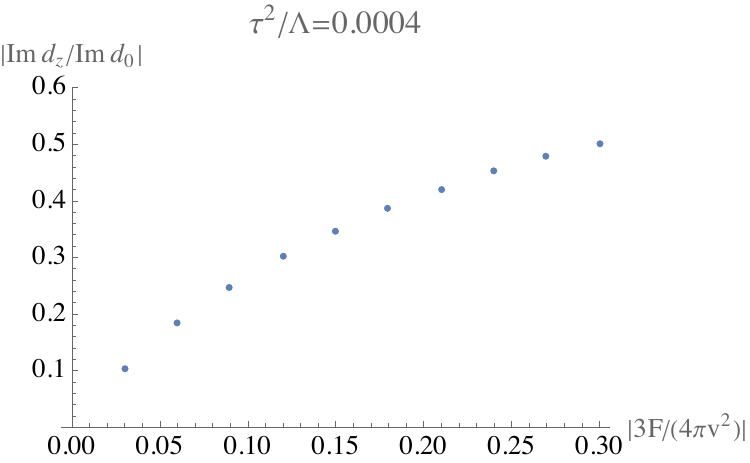}}
\caption{(a) The imaginary part of $\Sigma_{1}(0)$ as a function of different disorder strength square $F$ (unit $\omega^2/k^2$) from self-consistent Born$+$Dirac-cone approximation. (b) The ratio between the imaginary parts of $d$ vectors as a function of the disorder strength.}\label{fig_SCBAD}
\end{figure}

We also comment on the impact of the linear part ($\sim\omega$) in the self-energy in Eq.~\eqref{eq_snDSCBA}. At low frequencies, the linear term contributes a part $w\omega$ to the self-energy. This contribution can be eliminated by renormalizing all physical quantities related to orbital $1$. We should insert $H^{\mathrm{eff}}_{11}\to H^{\mathrm{eff}}_{11}/(1-w)$ and $H^{\mathrm{eff}}_{12}\to H^{\mathrm{eff}}_{12}/\sqrt{1-w}, H^{\mathrm{eff}}_{21}\to H^{\mathrm{eff}}_{21}/\sqrt{1-w}$. The eigenvalues of Green's function will be decided by this further modified effective Hamiltonian. Meanwhile, this renormalization also changes the weight of the quasiparticles at the corresponding poles of the Green's function. For small $w$, we can neglect such renormalization. But for a sufficient strong exceptional rings, $w$ is of the same order as $\textrm{Im }d_z/\textrm{Im }d_0$ and the renormalization is in principal not negligible. Especially, for the parameters taken in Fig~\ref{fig_SCBAD}, the linear term can suppress a non-negligible part of $\textrm{Im }d_z$. A more systematic way to study the effects of the linear term would be to treat the equation of motion for the Green's function as a generalized eigenvalue question. We will leave this to future study.

\subsection{The spin-spin correlation function}

In the presence of Heisenberg interactions $V_H=\sum_{\langle jk\rangle}\boldsymbol\sigma_j\cdot\boldsymbol\sigma_k$ and cross term $V_c=J_c\sum_{\alpha\beta\gamma}\sum_{\langle jk\rangle^\alpha}(\sigma_j^\beta\sigma_k^\gamma+\sigma_j^\gamma\sigma_k^\beta)$, the spin operator would be mixed with the low-energy Majoranas directly without exciting a flux. In this situation, there is an additional contributing to the spin operator, given by $\tilde\sigma_j^\alpha\sim \sum_{\beta\gamma}i\epsilon^{\alpha\beta\gamma}c_{j+\beta} c_{j+\gamma}$, where $c_{j+\beta}$ denotes the Majorana connected to $c_j$ via $\beta$ bond. For simplicity we look at $\tilde\sigma^z$. On the sublattice site $1$, it is given by $\tilde\sigma^z_j\sim ic_{j+\mathbf a_1}c_{j+\mathbf a_2}$ and on the sublattice site $2$ it's $\tilde\sigma^z_j\sim ic_{j-\mathbf a_1}c_{j-\mathbf a_2}$. In momentum space, this becomes $\tilde\sigma_{z}(\mathbf q)\sim \sum_{\mathbf k}ic_{2,\mathbf q-\mathbf k}c_{2,\mathbf k}\exp[i\mathbf q\cdot \mathbf a_1+i(\mathbf a_2-\mathbf a_1)\cdot \mathbf k]$ at site $1$ and $\tilde\sigma_{z}(\mathbf q)\sim \sum_{\mathbf k}ic_{1,\mathbf q-\mathbf k}c_{1,\mathbf k}\exp[-i\mathbf q\cdot \mathbf a_1+i(\mathbf a_1-\mathbf a_2)\cdot \mathbf k]$ at site $2$, where $\mathbf a_{1,2}$ are the lattice constants. As the Majoranas are not direct detectable, we need to consider the 
spin-spin correlation function $\chi_{zz}(t,\mathbf q)=-i\theta(t)\langle[ \tilde\sigma_z(t,\mathbf q),\tilde\sigma_z(0,-\mathbf q)]\rangle$. It has contributions from both the $\mathbb Z_2$ flux degrees of freedom and the Majorana degrees of freedom. At low energies, we only need to consider the Majorana part given by
\begin{equation}
    \chi^{(11)}_{zz}(t,\mathbf q)=\sum_\mathbf {k_1,k_2}i\theta(t)\left\langle[\psi_{2,-\mathbf k_1+\mathbf q}(t)\psi_{2,\mathbf k_1}(t),\psi_{2,-\mathbf k_2-\mathbf q}(0)\psi_{2,\mathbf k_2}(0)]\right\rangle e^{i(\mathbf a_2-\mathbf a_1)\cdot(\mathbf k_1+\mathbf k_2)}
\end{equation}
We use the following expression to translate the Green's function into the spectral function:
\begin{equation}
     G^{R}_{ab}(\omega, \mathbf k)=\int d\nu\frac{A_{ab}(\nu,\mathbf k)}{\omega-\nu+i0^+}, \quad \mathcal G_{ab}(i\omega, \mathbf k)=\int d\nu\frac{A_{ab}(\nu,\mathbf k)}{i\omega-\nu}.
\end{equation}
The spin correlation function is similar to a density-density correlation function for Majoranas and can be expressed through the Lindhard formula (here we neglect corrections brought by collective excitation for the underlying interacting system):
\begin{align}
    \chi^{(11)}_{zz}(\Omega, \mathbf q)\overset{\Omega+i0^+\to i\Omega}{=}
    &\sum_{\mathbf k,\omega}\mathcal G_{22}(i\omega+i\Omega, \mathbf k+\mathbf q)\mathcal G_{22}(i\omega, \mathbf k)(1-e^{i(\mathbf a_2-\mathbf a_1)\cdot(2\mathbf k+\mathbf q)})\nonumber\\
    \overset{i\Omega\to \Omega+i0^+}{=}&\sum_{\mathbf k} \int d\nu_1d\nu_2 \frac{n_F(\nu_1)-n_F(\nu_2)}{\nu_1-\nu_2-(\Omega+i0^+)}A_{22}(\nu_1,\mathbf k+\mathbf q)A_{22}(\nu_2,\mathbf k)(1-e^{i(\mathbf a_2-\mathbf a_1)\cdot(2\mathbf k+\mathbf q)}),\label{eq_ddksl}
\end{align}
where we omit a constant part of the density that only contributes at $\mathbf q=0$. In the above integral, when the frequency is continuously tuned, the variation in $\chi$ comes from a window of width $T$ near zero energy, $|\nu_1|,|\nu_2|\sim T$. A convenient to visualise this is to measure $\textrm{Im }\chi$. We need a bit techniques to separate this part out from \eqref{eq_ddksl}, where the phase factor $\exp[i(\mathbf a_2-\mathbf a_1)(2\mathbf k+\mathbf q)]$ can bring extra complications. The PH symmetry tells $A_{ab}(\nu,\mathbf k)=A_{ba}(-\nu,-\mathbf k)$. Using this property we can combine the two PH copies to make the phase factor $\exp[i(\mathbf a_2-\mathbf a_1)(2\mathbf k+\mathbf q)]$ real. So after doing replacement $\mathbf k\to\mathbf k-\mathbf q/2$, we have:
\begin{equation}
    \textrm{Im }\chi^{(11)}_{zz}(\Omega, \mathbf q)=\sum_{\mathbf k} \int d\nu\pi [n_F(\nu+\Omega)-n_F(\nu)]A_{22}\left(\nu+\Omega,\mathbf k+\frac{\mathbf q}{2} \right)A_{22}\left(\nu,\mathbf k-\frac{\mathbf q}{2}\right)\left[1-\cos[2(\mathbf a_2-\mathbf a_1)\cdot\mathbf k)]\right] .
\end{equation}
The factor $[n_F(\nu+\Omega)-n_F(\nu)]$ is only non-vanishing for $|\nu|<\Omega+T$. At zero temperature $T=0$, we can limit $-\Omega<\nu<0$ in the integral. For small $\Omega$, as the spectral function is now broadened by NH terms and becomes more continuous, we may further approximate 
\begin{equation}
    \textrm{Im }\chi^{(11)}_{zz}(\Omega, \mathbf q)\overset{\Omega\to 0}{\simeq} -\Omega\pi\sum_{\mathbf k} A_{22}\left(\frac{\Omega}{2},\mathbf k+\frac{\mathbf q}{2} \right)A_{22}\left(-\frac{\Omega}{2},\mathbf k-\frac{\mathbf q}{2}\right)\left[1-\cos[2(\mathbf a_2-\mathbf a_1)\cdot\mathbf k)]\right].
\end{equation}
This equation is non-vanishing for $\mathbf q$ connecting two points on nonzero regions of $A_{ab}(\omega,\mathbf k)$. 

Similarly, the the spin-spin spectral function at $22$ sublatice sites is given by 
\begin{equation}
    \textrm{Im }\chi^{(22)}_{zz}(\Omega, \mathbf q)=\sum_{\mathbf k} \int d\nu\pi [n_F(\nu+\Omega)-n_F(\nu)]A_{11}\left(\nu+\Omega,\mathbf k+\frac{\mathbf q}{2} \right)A_{11}\left(\nu,\mathbf k-\frac{\mathbf q}{2}\right)\left[1-\cos[2(\mathbf a_2-\mathbf a_1)\cdot\mathbf k)]\right]..
\end{equation}    
For the $12$ sublatice sites, the spin-spin correlation function itself is not Hermitian and we do a symmetrical combination. This makes it Hermitian and we can compute as below
\begin{align}
    \textrm{Im }[\chi^{(12)}_{zz}(\Omega, \mathbf q)+\chi^{(21)}_{zz}(\Omega, \mathbf q)]=&\sum_{\mathbf k} \int d\nu\pi [n_F(\nu+\Omega)-n_F(\nu)]\left[A_{21}\left(\nu+\Omega,\mathbf k+\frac{\mathbf q}{2} \right)A_{12}\left(\nu,\mathbf k-\frac{\mathbf q}{2}\right)e^{i(\mathbf a_2+\mathbf a_1)\cdot\mathbf q}\right.\nonumber\\
    &\times(e^{2i(\mathbf a_2-\mathbf a_1)\cdot\mathbf k}-1)+\textrm{c.c.}\Big].
\end{align}
The averaged spin-spin correlation function is given by summing over the above four components:
\begin{equation}
    \textrm{Im}\ \chi_{zz}(\Omega, \mathbf q)=\frac{1}{4}\sum_{ab} \textrm{Im}\ \chi^{(ab)}_{zz}(\Omega, \mathbf q).
\end{equation}
The spin-spin correlation function vanishes linearly as $\Omega\to0$ for small momentum. In this limit, we have
\begin{align}
     \frac{\textrm{Im}\ \chi_{zz}(\Omega, \mathbf q)}{\Omega}\simeq&-\frac{\pi}{4}\sum_{\mathbf k} \left\{\left[A_{22}\left(\frac{\Omega}{2},\mathbf k+\frac{\mathbf q}{2} \right)A_{22}\left(-\frac{\Omega}{2},\mathbf k-\frac{\mathbf q}{2}\right)+A_{11}\left(\frac{\Omega}{2},\mathbf k+\frac{\mathbf q}{2} \right)A_{11}\left(-\frac{\Omega}{2},\mathbf k-\frac{\mathbf q}{2}\right)\right][1\right.\nonumber\\&
     \left.-\cos[2(\mathbf a_2-\mathbf a_1)\cdot\mathbf k)]]
     +\left[A_{21}\left(\frac{\Omega}{2},\mathbf k+\frac{\mathbf q}{2} \right)A_{12}\left(-\frac{\Omega}{2},\mathbf k-\frac{\mathbf q}{2}\right)e^{i(\mathbf a_2+\mathbf a_1)\cdot\mathbf q}(e^{2i(\mathbf a_2-\mathbf a_1)\cdot\mathbf k}-1)+\textrm{c.c.}\right]\right\}
     .
\end{align}

According to the Lindhard formula \eqref{eq_ddksl}, the spin structure factor tells us for what total energy and momentum $(\Omega,\mathbf q)$ can a pair of quasiparticle and quasihole be excited. A effective dispersion of those quasiparticles is described by how peaks of $A_{ab}(\omega,\mathbf )$ in frequency $\omega$ moves with respect to the momentum $\mathbf k$.  In the Dirac fermion situation, at small $(\Omega,\mathbf q)$ we would expect that $\textrm{Im}\ \chi_{zz}(\Omega, \mathbf q)$ is only non-vanishing for $\Omega>v_Fq$ as shown in the main text. For the Fermi arcs, its contour in $A(\omega,\mathbf k)$ is biased and anisotropic at finite frequency (see figure in main text). This gives a very anisotropic effective dispersion and leads to the anisotropic result in the spin structure.

\subsection{Kitaev model with random magnetic field}
\label{app:random_field_model}

In this section we use perturbation theory to derive the Majorana model with second-neighbor bond disorder from the Kitaev honeycomb magnet with random magnetic field.
Following Ref.~\onlinecite{KITAEVHoneycomb}, let us add the term
\begin{equation}
V = - \sum_j \left(h^x_j \sigma^x_j + h^y_j \sigma^y_j + h^z_j \sigma^z_j \right)
\end{equation}
where $h^{\alpha}_j$ is the random variable corresponding to the $\alpha$ component of the magnetic field on site $j$.
In the following we consider zero mean, independent, isotropic random field, however, we allow the magnetic field fluctuations to be different on the $A$ and $B$ sublattices:
\begin{equation}
    {\rm Var}(h^{\alpha}_j) = \begin{cases}
\sigma^2_A \;\textnormal{if}\; j \in A,\\
\sigma^2_B \;\textnormal{if}\; j \in B.
\end{cases}
\end{equation}

We perform quasi-degenerate perturbation theory restricted to the vortex-free sector.
The first order correction vanishes, while the second order effective Hamiltonian contributes to the nearest-neighbor hoppings in the Majorana Hamiltonian
\begin{equation}
    H^{(2)}_{\rm eff} \approx \frac{i}{\Delta}\sum_{\langle j, k \rangle_{\alpha}} h^{\alpha}_j h^{\alpha}_k \sigma^{\alpha}_j \sigma^{\alpha}_k
    = \frac{i}{\Delta}\sum_{\langle j, k \rangle_{\alpha}} h^{\alpha}_j h^{\alpha}_k c_j c_k
\end{equation}
where $\langle j, k \rangle_{\alpha}$ runs over nearest neighbor bonds of $\alpha$ orientation.
Here we made the approximation of replacing the Green's function with the projector on the two-vortex sector divided by the gap to the state with two adjacent vortices $\Delta \approx 0.27 |J|$ in the isotropic case.
For the type of disorder we consider, this amounts to adding an uncorrelated (though not fully independent) noise with $u_{nn} = \sigma_A \sigma_B / \Delta$ standard deviation to the nearest neighbor hoppings.

At third order we find
\begin{equation}
    H^{(3)}_{\rm eff} \approx -\frac{1}{\Delta^2}\sum_{j, k, l} h^x_j h^y_k h^z_l \sigma^x_j \sigma^y_k \sigma^z_l
\end{equation}
where the sites $j, k, l$ are either three consecutive sites around a plaquette, or three nearest neighbors of a central site.
The second case corresponds to a four-majorana term and is irrelevant.
The first case results in a next-nearest-neighbor quadratic term
\begin{equation}
    H^{(3)}_{\rm eff} \approx -\frac{i}{\Delta^2}\sum_{\langle\langle j, k, l \rangle\rangle_{\alpha\beta\gamma}} h^{\alpha}_j h^{\beta}_k h^{\gamma}_l c_j c_l
\end{equation}
where $\langle\langle j, k, l \rangle\rangle_{\alpha\beta\gamma}$ corresponds to three sites such that $j$ and $k$ are connected by an $\alpha$ bond, $k$ and $l$ by a $\gamma$ bond and $\beta$ is different from both.
This results in uncorrelated random second neighbor hoppings with standard deviation $u_{AA} = \sigma_A^2 \sigma_B / \Delta^2$ for hoppings between $A$ sublattice sites and $u_{BB} = \sigma_A \sigma_B^2 / \Delta^2$ for hoppings between $B$ sublattice sites.

The resulting random variables are pairwise uncorrelated, but have higher order correlations.
In the numerical calculations we neglect these, and generate the nearest-neighbor noise from the product of two Gaussians drawn independently for every bond.
Similarly for the second neighbor noise we use products of three independent Gaussians.
As the self-consistent Born approximation is only sensitive to the second moments of the disorder distributions, the values of $F^{\bf l}$ are simply the variances of the hopping amplitudes on bonds corresponding to $\bf l$.

\subsection{Numerical results for disorder-averaged Green's functions}
\label{app:numerics}

We support the results obtained from self-consistent Born approximation with numerical calculations on finite disordered systems, finding good agreement between the two approaches.
In Fig.~\ref{fig:ring} we compare the single-particle spectral functions for a cut across the exceptional ring obtained from SCBA and disordered numerics.

All results were calculated for the isotropic model with $J_x = J_y = J_z = J = 1$, and we plot energies in units of $J$ throughout the manuscript.
For the figures in the main text and Figs.~\ref{fig:ring_comparison},~\ref{fig:ring_full_bz} we use random magnetic field $\sigma_A = J$ and $\sigma_B = (0.27/4) J = 0.0675 J$, which corresponds to nearest neighbor bond disorder with standard deviation $u_{nn} = 0.25 J$ and second neighbor bond disorders $u_{AA} \approx 0.926 J$, $u_{BB} = 0.0625 J$, and we use $\tau = {\rm Im}\;\omega = 0.02 J$.
To demonstrate that the exceptional rings persist for magnetic disorder that has magnitude smaller than the two-vortex gap, we repeated the calculation with $\sigma_A = J/2$ and $\tau = {\rm Im}\;\omega = 0.1 J$, all other parameters unchanged, shown in Fig.~\ref{fig:ring_small_disorder}.

We also compare the effective Hamiltonians $H_{\rm eff} = H_0 + \Sigma^R = \left(G^R\right)^{-1}$, where we evaluate the self-energy and the Green's function at ${\rm Re}\;\omega=0$.
The effective Hamiltonian becomes complex, with two complex eigenvalues $\tilde{E}_{\pm}$, and we plot the real and imaginary gaps in Fig.~\ref{fig:ring_comparison} for the above parameters.


\begin{figure}[t]
\includegraphics[width=0.5\linewidth]{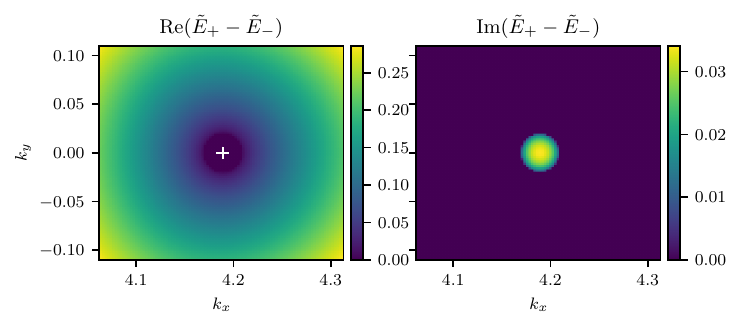}
\caption{Real (left) and imaginary (right) part of the gap obtained from self-consistent Born approximation with modified parameters $\sigma_A = J/2$ and $\tau = 0.1 J$.}
\label{fig:ring_small_disorder}
\end{figure}

\begin{figure}[t]
\includegraphics[width=0.5\linewidth]{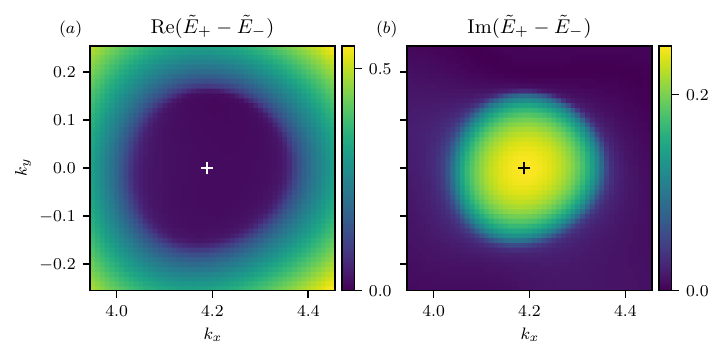}\\
\includegraphics[width=0.5\linewidth]{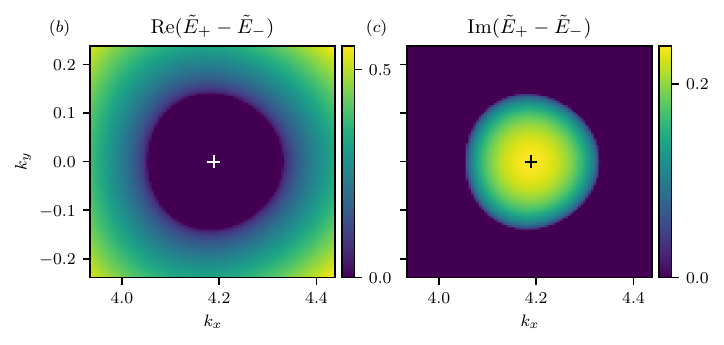}
\caption{Real ($(a)$ and $(c)$) and imaginary ($(b)$ and $(d)$) part of the gap in the effective Hamiltonian near the $K$ point obtained from disordered numerics ($(a)$ and $(b)$) and self-consistent Born approximation ($(c)$ and $(d)$).
The real gap vanishes inside, the imaginary gap outside the exceptional ring.}
\label{fig:ring_comparison}
\end{figure}

In the numerical calculations we use the Kwant software package~\cite{Groth2014} to construct the Hamiltonians for large disordered samples with periodic boundary conditions.
We use the Kernel Polynomial Method~\cite{Weisse2006, Varjas2019b} to approximate the retarded Green's function $G^R (\omega) = (\omega - H)^{-1}$ where ${\rm Im}\;\omega > 0$, and $H$ is the non-interacting tight-binding Majorana Hamiltonian written in real space for a finite sample with quasiperiodic boundary conditions.
This allows fast evaluation of matrix elements of the Green's function between normalized plane wave states $\ket{{\bf k}, a}$, which are only nonzero on the $a$ sublattice/orbital.
We use the self-averaging property, which states that taking such matrix elements in sufficiently large systems converges to the disorder-averaged Green's function:
\begin{equation}
    \langle G^R_{ab} (\omega, {\bf k}) \rangle_{\rm dis} \approx  \bra{{\bf k}, a} G^R (\omega) \ket{{\bf k}, b}.
\end{equation}
For the numerical results shown here we achieve self-averaging by using systems with $2 \times 10^4$ sites.

We calculate the density of states (DoS) from the real space Green's function as $\rho(\omega) = -\frac{1}{\pi} \operatorname{Im\;Tr} G^R(\omega)$.
We use Fermi liquid theory to obtain the specific heat, setting the chemical potential to zero:
\begin{equation}
    c = \frac{du}{dT} = \frac{d}{dT}\int_{-\infty}^{\infty} \omega \rho(\omega) f(T, \omega) d\omega,
\end{equation}
where $f$ is the Fermi-Dirac distribution.
Using the Sommerfeld expansion, and using the fact that particle-hole symmetry forces $\rho$ to be an even function, we find
\begin{equation}
    \frac{c}{k_B} = \frac{\pi^2}{3} \rho(0) \;k_B T + \frac{7 \pi^4}{30}\frac{d^2\rho}{d\omega^2}(0)\; (k_B T)^3 + \mathcal{O}\left((k_B T)^5\right),
\end{equation}
showing that the sign of the lowest order (cubic) deviation from the linear temperature dependence of the specific heat is determined by the second derivative of the DoS at zero frequency.
Note that this calculation assumes that the DoS is analytic at $\omega=0$, which is not satisfied in the clean Dirac cone limit where $\rho(\omega)\propto |\omega|$, resulting in a quadratic specific heat at lowest order.

\begin{figure}[t]
\includegraphics[width=0.95\linewidth]{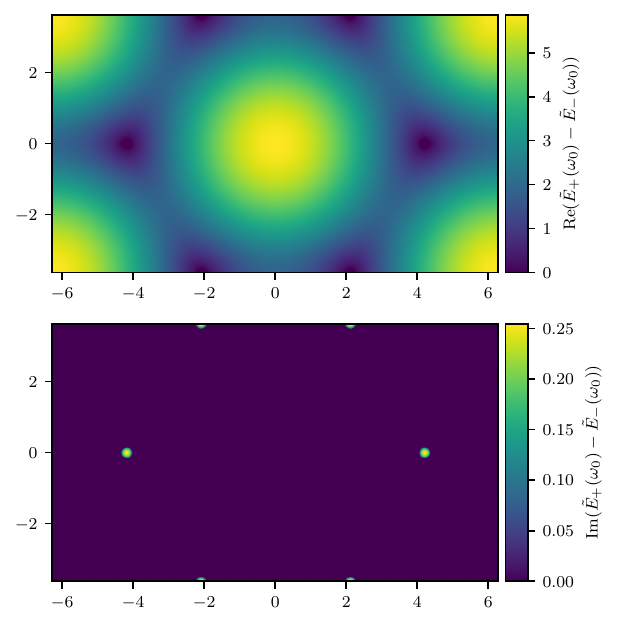}
\caption{Real (top) and imaginary (bottom) part of the gap in the effective Hamiltonian with $\tau = 0.02$ obtained from self-consistent Born approximation.
The exceptional rings are visible at the $K$ and $K'$ points of the Brillouin zone.}
\label{fig:ring_full_bz}
\end{figure}


\end{widetext}

\end{document}